# Valley polarization by spin injection in a light-emitting van der Waals heterojunction

Oriol Lopez Sanchez[1,2], Dmitry Ovchinnikov[1,2], Shikhar Misra[3], Adrien Allain[1,2],
Andras Kis[1,2*]

[1]Electrical Engineering Institute, École Polytechnique Fédérale de Lausanne (EPFL), CH-1015 Lausanne, Switzerland
[2]Institute of Materials Science and Engineering, École Polytechnique Fédérale de Lausanne (EPFL),
CH-1015 Lausanne, Switzerland
[3]Department of Materials Science and Engineering, Indian Institute of Technology (IIT), Kanpur 208016, India
*E-mail address: andras.kis@epfl.ch

The band structure of transition metal dichalcogenides (TMDCs) with valence band edges at different locations in the momentum space could be harnessed to build devices that operate relying on the valley degree of freedom. To realize such valleytronic devices, it is necessary to control and manipulate the charge density in these valleys, resulting in valley polarization. While this has been demonstrated using optical excitation, generation of valley polarization in electronic devices without optical excitation remains difficult. Here, we demonstrate spin injection from a ferromagnetic electrode into a heterojunction based on monolayers of $WSe_2$ and $MoS_2$ and lateral transport of spin-polarized holes within the $WSe_2$ layer. The resulting valley polarization leads to circularly polarized light emission which can be tuned using an external magnetic field. This demonstration of spin injection and magnetoelectronic control over valley polarization provides a new opportunity for realizing combined spin and valleytronic devices based on spin-valley locking in semiconducting TMDCs.

While most electronic devices are based on the manipulation of the electric charge, alternative schemes based on the manipulation of spins in spintronic devices[1] or valleys in valleytronic devices are desirable. The use of the valley index for information processing was initially proposed using AlAs and Si[2] and more recently in the case of graphene.[3] Ultrathin transition metal dichalcogenides with the common formula $MX_2$ ($MoS_2$, $MoSe_2$, $WS_2$, $WSe_2$,..)[4] are emerging as an interesting material system for the realization of valleytronic devices. Field-effect transistors[5] and optoelectronic devices[6–11] have already been realized using semiconducting TMDCs which are direct band gap semiconductors in the monolayer limit.[12–15]

Inversion symmetry breaking in TMDC monolayers and a large mass of constituent atoms results in strong spin-orbit coupling and the splitting of the valence band of $MX_2$ monolayers into two subbands with spin up and spin-down states,[16,17] with the splitting having opposite signs for the two inequivalent valleys (K and K´). Previous theoretical studies predicted a valence band splitting between 0.15 eV for $MoS_2$ to 0.46 eV for $WSe_2$ composed of heavier atoms.[18,19] More recent models also predict a ~40 meV conduction band splitting for $WS_2$ and $WSe_2$.[18] The valley index is expected to be protected from scattering by smooth deformations and long-wavelength phonons because intervalley mixing would require simultaneous spin flipping and scattering from phonons.[17,20] This gives the possibility of using the valley index as the information carrier in valleytronic devices.

Establishing valley polarization using optical excitation with circularly polarized light[21,22] has already been demonstrated while the valley state can also be electrically detected *via* the valley Hall effect.[17,23] However, achieving and controlling the valley polarization without the use of optical excitation,[24] desirable for realizing complex valleytronic devices and circuits, remains difficult. Since charge carriers at band edges in the K and K´ valleys in TMDCs carry opposite spins, this spin-valley locking[17] gives the opportunity to interchangeably address the spin and valley degrees of freedom and to achieve valley polarization by injecting spin-polarized charge carriers into TMDC materials. Alternative approaches involve possibly exploiting the anisotropy of the Fermi pockets and/or electron-hole overlap in 2D TMDCs,[24] either upon current injection[11] or using ac electric fields.[24]

Even though valley polarization by spin injection can be realized in a vertical geometry by using a p-type magnetic semiconductor Ga(Mn)As,[25] the use of a magnetic substrate for spin injection limits possible device geometries and further development of integrated valleytronic devices and circuits. For this, spin injection using local electrodes and spin transport in lateral direction needs to be achieved.

We demonstrate here that spin-polarized charge carriers can indeed be injected into monolayer $WSe_2$ using local permalloy electrodes and transported inside the 2D semiconductor in lateral direction. We realize a light-emitting diode to demonstrate spin injection[26,27] by integrating $WSe_2$ with $MoS_2$ to form a vertical p-n junction,[28–30] capable of electroluminescence.[29] This geometry allows us to keep the device simple, since we do not need to use additional local gate electrodes to induce a lateral p-n junction.[7–11] We can control the circular polarization of emitted light by injecting spin polarized carriers with a permalloy electrode. On Fig. 1a we show an optical micrograph of our device based on a van der Waals heterostructure composed of monolayers of MoS$_2$ and WSe$_2$, forming an atomically sharp vertical p-n junction.[28–30] It is assembled by transferring MoS$_2$ on top of WSe$_2$ exfoliated on a SiO$_2$ substrate. Electrical contacts to both 2D semiconductors are fabricated using electron-beam lithography. We show the band alignment diagram for our device on Supplementary Figure 1. Au is used for injecting electrons into MoS$_2$,[5,31] while WSe$_2$ is contacted using ferromagnetic permalloy (Ni:Fe alloy in a 81:19 ratio, 75 nm thickness) with the aim of injecting spin-polarized holes into the heterostructure, Fig. 1b. The Fermi energy of permalloy ($E_F$ = 4.8 eV)[32] is very close to the calculated top of the valence band in monolayer WSe$_2$ ($E_V$ = 4.86 eV),[33] making this material suitable for hole injection into WSe$_2$. A 30 nm thick HfO$_2$ layer is deposited over the entire heterojunction in order to encapsulate the device and improve its stability. In order to verify spin injection in our heterostructures, we also fabricate a set of non-magnetic devices, with a Pd/Au electrode in place of permalloy.

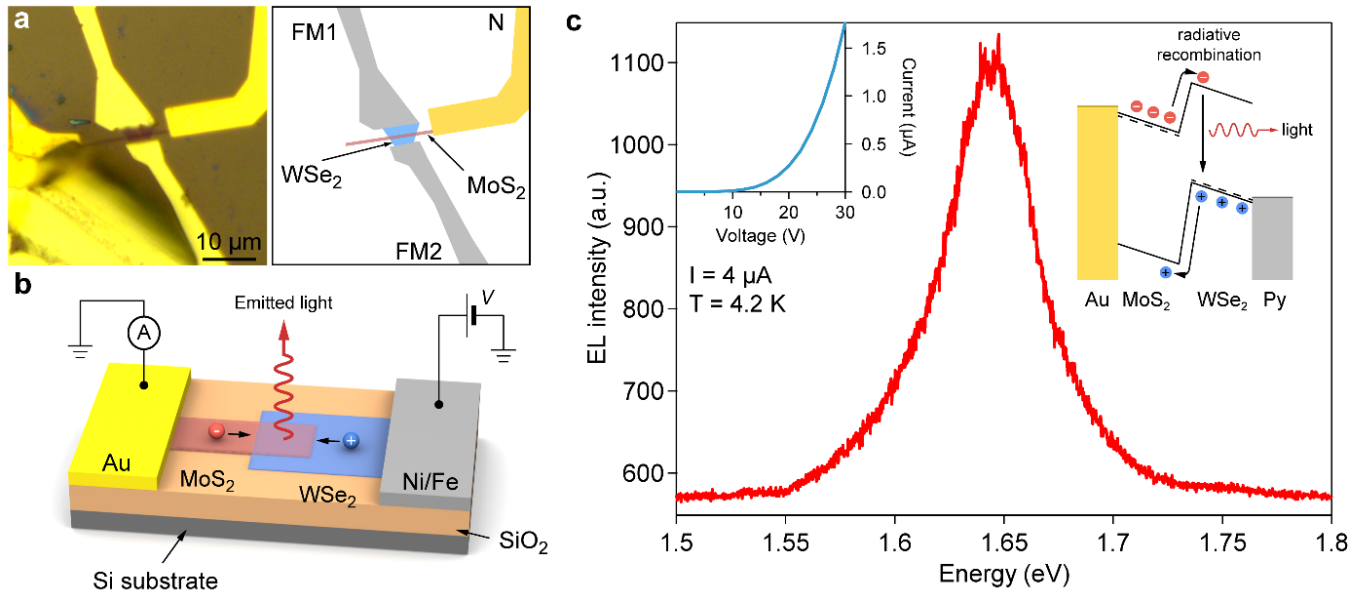

Fig. 1 Light generation in monolayer WS$_2$/MoS$_2$ heterojunction diodes. **a,** Optical image of the device. Monolayer WSe$_2$ is contacted using a ferromagnetic electrode (permalloy). MoS$_2$ is transferred on top of the MoS$_2$ channel, forming a heterojunction diode. **b,** Schematic drawing of the device. Under the application of a positive bias voltage to the permalloy electrode, holes are injected from the permalloy electrode and recombine in the junction with electrons injected from the MoS$_2$ side, resulting in light emission. **c,** Electroluminescence spectrum, showing light emission dominated by the X$^-$ trion resonance. Insets: current-voltage characteristic of the device and the band diagram of the device under forward bias.

Under the application of a positive bias voltage to the permalloy electrode, holes are injected from it into WSe$_2$, while electrons are injected from the gold electrode into MoS$_2$, Fig. 1b. At forward bias, the heterojunction p-n diode can emit light, with the highest emitted light intensity at the edge of the heterojunction (Supplementary Fig. 3). Fig. 1c shows a representative electroluminescence (EL) spectrum acquired from the device at 4.2 K for a forward current of 4 µA. We do not observe any light emission or appreciable current for a reverse bias (Supplementary Figure 3) which is indicative of a pn junction. The EL spectrum is related to the direct band gap of WSe$_2$[10] and not MoS$_2$ or the interlayer band gap of the MoS$_2$/WSe$_2$ heterostructure,[28] indicating that the recombination process is governed by the band structure of WSe$_2$. This is due to the band alignment[33] between MoS$_2$ and WSe$_2$ as well as the smaller band gap of WSe$_2$ which results in conduction band offset of $\Delta E_C$ = 0.67 eV and a valence band offset of $\Delta E_V$ = 1 eV. Because of this, the injection barriers for holes from WSe$_2$ into MoS$_2$ are higher than for electrons from MoS$_2$ into WSe$_2$ and the recombination is more likely to occur in WSe$_2$.

We assign the dominant feature to the negatively charged X$^-$ trion,[10] composed of two electrons and a hole,[10] due to an excess of electrons because of n-type doping induced by the HfO$_2$ passivation layer.[31] At low injected current densities, the impurity bound exciton[10] X$^I$ and the neutral exciton X$^0$ can also be resolved. Their relative contributions however decrease as the injected current is increased (Supplementary Fig. 2). Previous polarization-dependent photoluminescence studies have shown that valley polarization can be optically induced in monolayer WSe$_2$ with excitons being formed in K and K′ valleys.[34] Because the holes at the valence band edges at the two valleys have opposite spins,[17] using a ferromagnetic electrode to inject spin polarized holes into WSe$_2$ allows us to break the valley symmetry and simultaneously induce spin and valley polarization without the use of polarized optical excitation. The optically active heterojunction then serves as the spin-valley detector. In graphene-based devices, the high mobility of the material and low contact resistances require using tunnel barriers at contacts[35] in order to achieve efficient spin injection and to mitigate the impedance matching problem encountered in the realization in semiconductor-based lateral spin valves.[36,37] In the case of TMDC materials, the relatively high contact resistance could be a solution to the problem of conductivity mismatch and the already present Schottky barriers could allow for efficient spin injection[38,39] without artificially introducing tunnel barriers, thus simplifying device fabrication. Instead of fabricating a spin valve, we use a light-emitting heterostructure. It presents a convenient way of demonstrating spin injection into semiconductors,[26,27] because the detection mechanism does not require a second, less-than ideal ferromagnetic electrode like in the case of spin valves which would result in the accumulation of losses in spin injection, transport and detection. On Fig. 3 we illustrate the principle of valley polarization detection in the electroluminescent device. Fig. 3a shows the four configurations of X⁻ charged exciton that can emit light, with two degenerate sets due to electron-hole exchange.[40] Optical transitions are allowed for electron-hole pairs in the same valley, resulting in the emission of circularly polarized light and an extra electron. Spin-polarized holes can be injected into $WSe_2$ from the permalloy electrode, Fig. 3b. Since the valence band energy is ~0.5 eV lower at the Γ point than at the K-K´ points[16] of monolayer $WSe_2$, spin injection occurs at the K-K´ points. Spin polarized holes recombine in the heterojunction area with unpolarized electrons from $MoS_2$. Because the electronic states in the K and K´ valleys have different chiralities due to breaking of inversion symmetry, interband transitions at band edges involve $\sigma_+$ and $\sigma_-$ polarized light. Using an external magnetic field, we can control the magnitude and orientation of the magnetization of the permalloy electrode. Since the spins at the valence band edge in K and K´ valleys are aligned in the direction perpendicular to the 2D sheet,[16] the magnetic field is applied in a direction normal to the surface of the heterojunction (Faraday geometry) and swept in the ±5T range. Due to spin-valley locking, spin polarization of holes at band edges results in their valley polarization which leads to different intensities in the emission of σ$_+$ and σ$_-$ polarized light from the heterojunction as a result of valley polarisation. The external magnetic field also modulates the transition energy by $\Delta = \Delta E_C - \Delta E_V$ via the valley Zeeman effect[41,42] which shifts the conduction and valence band edges by $\Delta E_C$ and $\Delta E_V$ respectively.

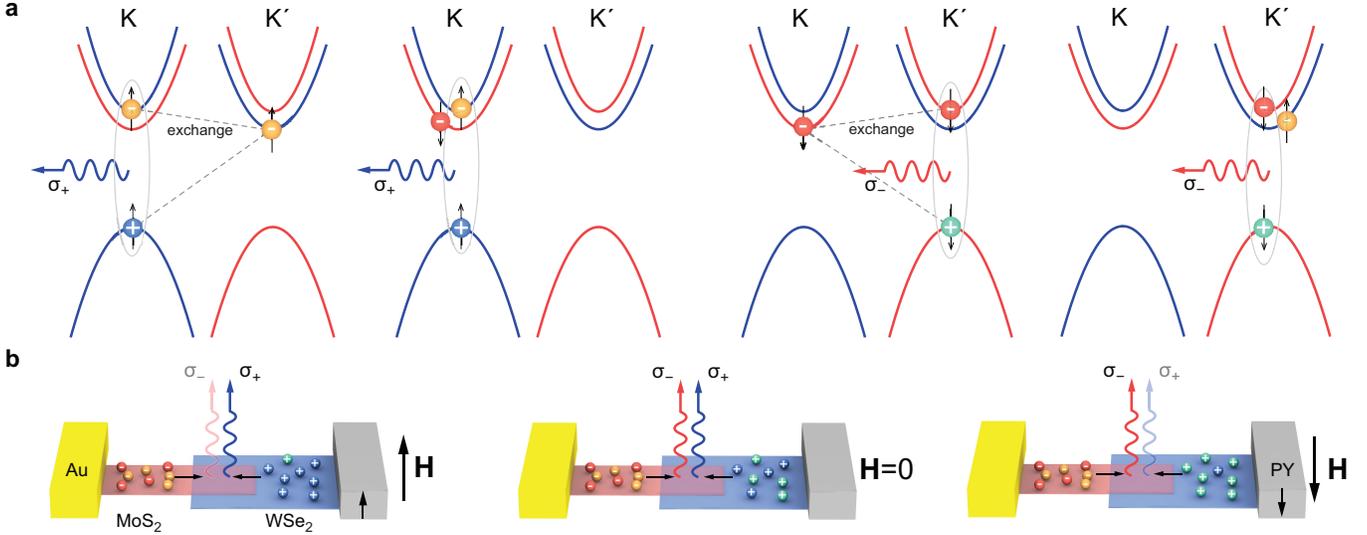

Fig. 3 Principle of valley polarization by spin injection and its detection in a spin-valley LED. **a**, Configurations of X⁻ charged excitons in $WSe_2$ that can emit light. Blue and red lines denote spin up and spin down polarized conductance and valence band edges in the K and K´ valleys. The polarization of the emitted photon is determined by the valley index of the electron-hole pair that can recombine. **b**, Application of an external magnetic field results in the magnetization of the permalloy electrode and injection of spin-polarized holes. In a positive (negative) external magnetic field, a majority of spin up (spin down) holes is injected on the $WSe_2$ side of the device. This results in valley symmetry breaking and enhanced emission of $\sigma_+$ ($\sigma_-$) polarized light.

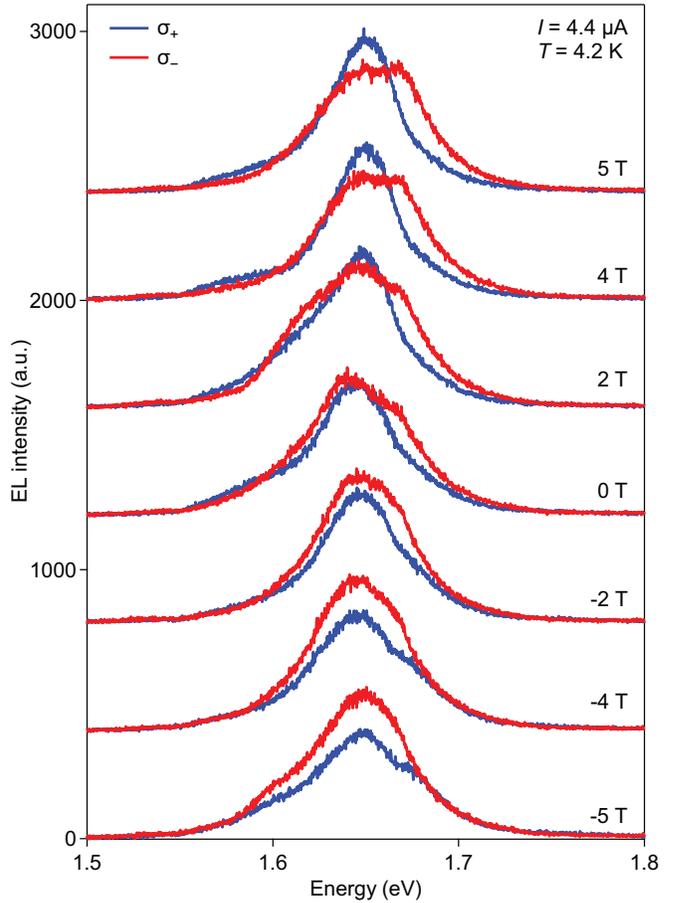

Fig. 2 Electroluminescence spectra from a spin/valley LED for different values of magnetic fields, acquired for $\sigma_+$ and $\sigma_-$ polarizations.

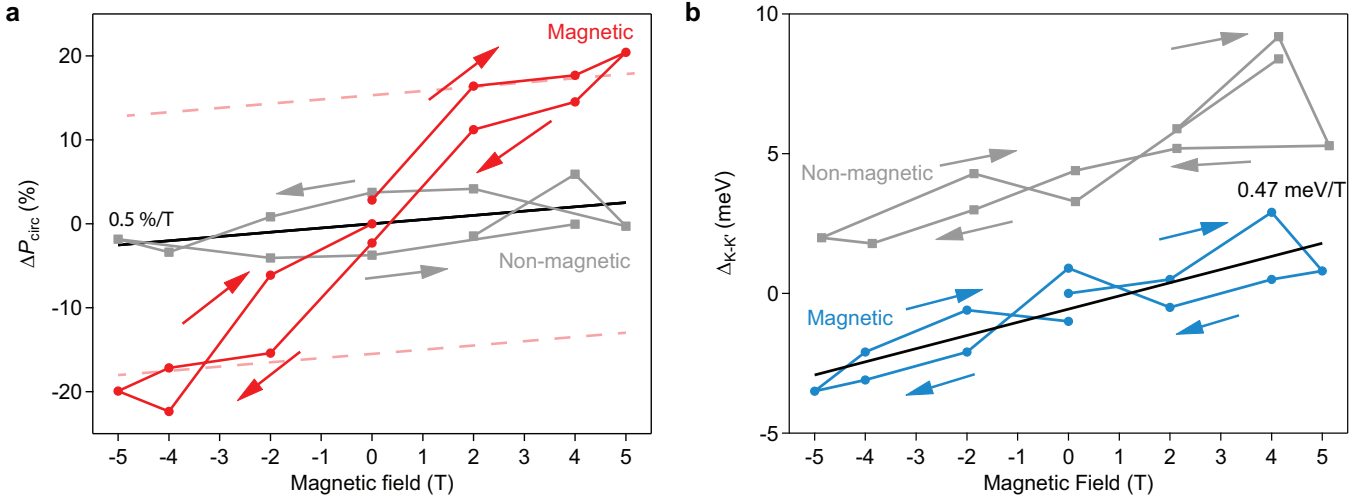

Fig. 4 Magnetic field dependence of electroluminescence polarization and the valley Zeeman effect in a WSe$_2$/MoS$_2$ heterojunction. **a**, Dependence of the relative difference in degree of circular polarization of the X$^-$ resonance as a function of external magnetic field for a device with a magnetic (PY) electrode and for a control device with non-magnetic electrodes. Dashed red lines run parallel to the ~0.5 %/T contribution to valley polarization in WSe$_2$ due to the applied magnetic field. **b**, Valley Zeeman effect for the X$^-$ resonance detected in electroluminescence for the device with magnetic electrodes and for the control device with non-magnetic electrodes. Results for the control device are offset in vertical direction for clarity.

EL spectra for different values of the external magnetic field acquired from the device at 4.2 K are shown on Fig. 2. The device is driven at an injection current $I$ = 4.4 µA and the resulting electroluminescence spectra are acquired simultaneously for the σ$_+$ and σ$_-$ circular polarizations with the use of a calcite beam displacer[22] which excludes possible artifacts that could be related to time-dependent fluctuations in the device output. At zero field, σ$_+$ and σ$_-$ components are overlapping, with a small difference of ~3% due to the alignment accuracy of the calcite beam displacer. As the field is increased, the permalloy contact is magnetized and is injecting spin-polarized holes into the heterostructure. The spin-imbalance of holes breaks the valley symmetry resulting in the emission of circularly polarized light. At −5T, the emission of light with σ$_+$ polarization, corresponding to interband transitions in the K valley, is suppressed, while EL for σ$_-$ (transitions in the K′ valley) it is enhanced. As the field direction is reversed, emission of σ$_+$ polarized light is enhanced and σ$_-$ suppressed. This is the first demonstration of spin injection and resulting valley polarization in a heterostructure based on 2D semiconductors.

We extract the magnitude of spin polarization from the EL spectra by fitting them to three Lorentzian curves, corresponding to X$^-$, X$^0$ and X$^I$ transitions, Supplementary Figure 4. We concentrate on the dominant peak, attributed to the X$^-$ charged exciton and neglect other features in the spectra since the extraction of their intensities would be much more strongly dependent on the fit quality than in the case of the dominant X$^-$ transition. We define the degree of circular polarization for the X$^-$ transition as $P_{circ} = (I_+ - I_-)/(I_+ + I_-)$ where $I_+$ and $I_-$ correspond to peak intensities of the X$^-$ exciton for the σ$_+$ and σ$_-$ polarizations. On Fig. 4a we show the magnetic field dependence of circular polarization for the X$^-$ component by plotting $\Delta P_{circ}(B) = P_{circ}(B) - P_{circ}(B = 0)$, where $P_{circ}(B = 0) = 3\%$ is the systematic error due to the alignment accuracy of the calcite beam displacer. Because the external magnetic field is aligned along the hard magnetic axis of the 75 nm thick permalloy electrode, the magnetic response of circular polarization shows only a very small amount of hysteresis.

In addition to valley polarization due to the injection of spin-polarized holes, the circular polarization of the light emitted by our devices with a ferromagnetic contact also contains a contribution due to valley polarization in WSe$_2$ induced by the external magnetic field.[43,41] In order to distinguish between these two contributions, we also fabricate a device with non-magnetic electrodes (Supplementary section 4). In contrast to the device with permalloy electrodes, the control device shows a much smaller change in the degree of circular polarization as the external magnetic field is swept (Fig. 4a and Supplementary Figure 6), on the order of ~0.5%/T. For a magnetic field of 5 T, we reach a total $\Delta P_{circ} \approx 20\%$. After subtraction of the contribution from the valley polarization in WSe$_2$ due to the magnetic field of $P_{in} \approx 2.5\%$ at 5 T, the contribution to circular polarization due to the spin polarization of charge carriers $P_{spin} = P_{circ} - P_{in} \approx 17.5\%$ (Supplementary Figure 7), showing that the permalloy electrode can efficiently inject spin-polarized charge carriers. Future contact optimization could result in even higher levels of spin polarization.

While the injection of spin-polarized charge carriers from the magnetized electrode can induce valley polarization i.e. cause an imbalance in the charge distributions in K and K′ valleys, the application of a magnetic field in the Faraday geometry also lifts the degeneracy via the valley Zeeman effect,[41,42] previously seen in photoluminescence experiments on TMDC monolayers. Here, we also observe the valley Zeeman effect *via* electroluminescence and in heterostructures. It gives rise to splitting between the X⁻ transition peaks in σ₊ and σ₋ electroluminescence spectra. Fig. 4b shows the magnetic field dependence of the splitting $\Delta_{K-K'}$ after subtraction of $\Delta_{K-K'}(B = 0) = 4.2$ meV due the alignment accuracy of the calcite displacer. A slope of $0.47 \pm 0.08$ meV/T is found, reflecting high charge densities in our device due to doping induced by the deposition of the $HfO_2$ passivation layer. Since the valley Zeeman effect is due to a difference in magnetic-field induced shifts of the conduction and valence bands in the TMDC monolayer and does not require the injection of spin-polarized charge carriers, the control device with non-magnetic electrodes shows a similar valley Zeeman effect as the device with the permalloy electrode, Fig. 4b.

We have used a light-emitting diode based on the van der Waals heterostructure geometry to demonstrate spin injection and resulting valley polarization in a 2D semiconductor. Circularly polarized electroluminescence from the LED shows that spin-polarized charge carriers can be injected across the Schottky barrier between a ferromagnetic electrode and $WSe_2$ and transported inside the 2D semiconductor, resulting in valley polarization due to spin-valley locking in TMDC materials and in a device geometry that can enable more complicated structures in future. The degree of spin/valley polarization can be manipulated by controlling the magnetization of the injecting electrode. Our demonstration of spin injection into a 2D-semiconductor based heterostructure and a spin-valley LED tunable via an external magnetic field proposes a way to control the valley polarization in TMDC materials without optical excitation and could allow a host of new valleytronic devices based on electrical and magnetic control of valley polarization in 2D semiconductors.

## MATERIALS AND METHODS

Single layers of $MoS_2$ (SPI supplies) and $WSe_2$ (2Dsemiconductors.com) are exfoliated from bulk crystals onto silicon substrates covered by a 270 nm thick layer of thermal oxide. Monolayer samples are identified by optical microscopy[44]. Once identified, monolayer $MoS_2$ is transferred onto $WSe_2$ using a polymer stamp transfer technique.[28] Electrical contacts are fabricated using e-beam lithography and thermal evaporation of 75 nm thick permalloy, capped with 25 nm of gold in order to prevent oxidation in air. Light-emitting diodes are characterized at 4.2 K in an optical cryostat with a built-in superconducting magnet (Oxford Instruments). The emitted radiation was collected and analyzed using a grating spectrometer (Andor Shamrock 500i) equipped with a thermoelectrically cooled CCD camera (Andor Newton 970). Electrical measurements were performed using a DAQ card (National Instruments NI-PXI 4461) and a current amplifier (Stanford Research SR570). Spectra for σ₊ and σ₋ polarizations are simultaneously acquired using a calcite beam displacing prism[22].

## ACKNOWLEDGEMENTS

Device fabrication was carried out in the EPFL Center for Micro/Nanotechnology (CMI). We thank Z. Benes (CMI) for technical support with e-beam lithography. This work was financially supported by the Swiss National Science Foundation (grant no. 153298) and European Research Council (ERC, grant no. 682332) and was carried out in frames of the Marie Curie ITN network "MoWSeS" (grant no. 317451). We acknowledge funding by the EC under the Graphene Flagship (grant agreement no. 604391).

## SUPPORTING INFORMATION AVAILABLE

Supplementary figures and discussion related to material band alignment, electroluminescence spectra evolution with the magnetic field, electroluminescence mapping, emission curve fitting and the behavior of control devices with non-magnetic electrodes. This material is available free of charge via the Internet at http://pubs.acs.org.